%% file: Proceedings_Husung.tex
\documentclass{PoS}
\pdfoutput=1

\usepackage{amssymb,mathrsfs,mathtools,slashed}

\title{Logarithmic corrections to $\mathbf{a^2}$ scaling in lattice Yang Mills theory}

\ShortTitle{Logarithmic corrections to $a^2$ scaling in lattice Yang Mills theory}

\author{\speaker{Nikolai Husung}$^a$, Peter Marquard$^b$, and Rainer Sommer$^{a,c}$\\
\llap{$^a$}John von Neumann Institute for Computing (NIC), DESY, Platanenallee 6, 15738 Zeuthen, Germany\\
\llap{$^b$}Theory group, DESY, Platanenallee 6, 15738 Zeuthen, Germany\\
\llap{$^c$}Institut f\"ur Physik, Humboldt-Universit\"at zu Berlin, Newtonstr. 15, 12489 Berlin, Germany\\
E-mail: \email{nikolai.husung@desy.de}, \email{peter.marquard@desy.de}, \email{rainer.sommer@desy.de}}

\abstract{We analyse the leading logarithmic corrections to the $a^2$ scaling of lattice artefacts in QCD, following the seminal work of Balog, Niedermayer and Weisz in the O(n) non-linear sigma model.
Restricting our attention to contributions from the action, the leading logarithmic corrections can be determined by the anomalous dimensions of a minimal on-shell basis of mass-dimension 6 operators.
We present results for the SU(N) pure gauge theory.
In this theory the logarithmic corrections reduce the cutoff effects.
These computations are the first step towards a study of full lattice QCD at O($a^2$), which is in progress.
\vspace{0.5cm}
\begin{flushright}
  DESY 19-200
\end{flushright}}

\FullConference{37th International Symposium on Lattice Field Theory - Lattice2019\\
		16-22 June 2019\\
		Wuhan, China}

\def\ord{\mathrm{O}}
\newcommand{\tr}{\mathrm{tr}}
\newcommand{\diag}{\mathop{\mathrm{diag}}}
\newcommand{\Pexp}{\mathop{\mathrm{Pexp}}}

\newcommand{\Exp}[1]{\left\langle #1\right\rangle}

\newcommand{\lat}[2][]{{#2}_{\ifx \EMPTY#1\EMPTY \else #1;\fi\mathrm{lat}}}
\newcommand{\R}[2][]{{#2}_{\ifx \EMPTY#1\EMPTY \else #1;\fi\mathrm{R}}}
\newcommand{\RGI}[2][]{{#2}_{#1\mathrm{RGI}}}
\newcommand{\bare}[2][]{{#2}_{\ifx \EMPTY#1\EMPTY \else #1;\fi 0}}
\newcommand{\lattExp}[2][\mathrm{R}]{\Exp{#2}_{\substack{\mathrm{con}\hfill\\\mathrm{lattice}\ifx \EMPTY#1\EMPTY \else ;#1\fi\hfill}}}
\newcommand{\contExp}[2][\mathrm{R}]{\Exp{#2}_{\substack{\mathrm{con}\hfill\\\mathrm{cont}\ifx \EMPTY#1\EMPTY \else ;#1\fi\hfill}}}
\newcommand{\latt}[2][\mathrm{R}]{#2_{\mathrm{latt}(a)\ifx \EMPTY#1\EMPTY \else ;#1\fi}}
\newcommand{\cont}[2][\mathrm{R}]{#2_{\mathrm{cont}\ifx \EMPTY#1\EMPTY \else ;#1\fi}}

\renewcommand{\L}{\mathscr{L}}
\newcommand{\Leff}{\L_\mathrm{eff}}

\def\lcf{\Phi}
\def\lcfr{\Phi_\mathrm{R}}
\def\lcfrlat{\Phi_\mathrm{lat}}
\def\dlcf{\Upsilon}
\def\dlcfr#1{\Upsilon_{#1;\mathrm{R}}}
\def\dummy{\chi}
\newcommand{\op}{\mathcal{O}}
\newcommand{\E}{\mathcal{E}}
\newcommand{\opT}{\slashed{\op}}
\newcommand{\basis}{\mathcal{B}}

\def\CA{C_\mathrm{A}}

\newcommand{\EOM}[1][=]{\stackrel{\text{\tiny EOM}}{#1}}
\newcommand{\D}{\text{d}}

\newcommand{\TL}[1]{#1^{(0)}}

\def\bop{c}
\def\bbase{\bar{c}}

\def\cop{d}

\def\RI{\mathrm{RI}}
\def\xRI{x_\RI}

\let\OLDthebibliography\thebibliography
\renewcommand\thebibliography[1]{
  \OLDthebibliography{#1}
  \setlength{\parskip}{1pt}
  \setlength{\itemsep}{0.5pt plus 0.3ex}
}

\begin{document}

\section{Introduction}
Symanzik effective field theory~\cite{Symanzik:1979ph,Symanzik:1981hc,Symanzik:1983dc,Symanzik:1983gh} has been used extensively for so called on-shell improvement.
This removes the lattice artifacts at leading order in the lattice spacing~$a$~\cite{Sheikholeslami:1985ij,Weisz:1982zw,Weisz:1983bn,Luscher:1985}.
We follow here a related path and compute the leading logarithmic corrections to classical $a^2$ scaling in pure gauge theory originating from the action as a first step towards full lattice QCD.
% Note that $\ord(a^2)$ is the leading order for unimproved pure gauge theory.
This fills a gap in the theoretical control over  continuum extrapolations, where one usually assumes pure $a^2$~scaling, although for an asymptotically free theory leading corrections have the form $a^2[\bar{g}^2(1/a)]^{\hat{\gamma}_1}$ with coupling $\bar{g}^2(1/a)\sim -1/\ln(a\Lambda)$ for small $a$. The
parameter $\hat{\gamma}_1$ is proportional to a 1-loop anomalous dimension.
Therefore, whether a continuum extrapolation with classical $a^2$ scaling is accurate, depends on $\hat{\gamma}_1$ and the available lattice spacings.

For example in the 2d non-linear sigma-model, classical $a^2$ scaling turned out far from realistic.
There a pioneering analysis has been performed~\cite{Balog:2009yj,Balog:2009np} solving a long-standing puzzle of lattice artifacts, which seemed to behave more like $\ord(a)$ than the classically expected $\ord(a^2)$. The issue was solved by large logarithmic corrections, i.e. distinctly negative $\hat{\gamma}_1$.

\section{Symanzik effective theory}\label{sec:SET}
To describe lattice artifacts impacting expectation values obtained on lattices with finite lattice spacing we use a continuum Symanzik effective theory~\cite{Symanzik:1979ph,Symanzik:1981hc,Symanzik:1983dc,Symanzik:1983gh}.
There are different sources of lattice artifacts.
Firstly, the discretised action described by the effective Lagrangian
\begin{equation}
\Leff=\L+a^2\delta\L+\ord(a^4),
\end{equation}
with the Lagrangian of the targeted continuum theory
\begin{equation}
\L=-\frac{1}{2g_0^2}\tr(F_{\mu\nu}F_{\mu\nu}),\quad F_{\mu\nu}=\partial_\mu A_\nu-\partial_\nu A_\mu+[A_\mu,A_\nu]
\end{equation}
and secondly any local field $\lcfr$ involved in correlation functions of interest\footnote{
Renormalised local fields have to be used to match the lattice theory and the effective theory.}
\begin{equation}
\lcf_\mathrm{eff;R}(x)=\lcfr(x)+a^2\delta\lcf(x)+\ord(a^4).
\end{equation}
Thirdly, the renormalisation condition employed on the lattice.
For now we will assume the use of lattice perturbation theory in the MS lattice~(lat) renormalisation scheme~\cite{Luscher:1995np} and postpone the inclusion of the third contribution to the end of this section.

Consequently, to parametrise all possible lattice artifacts in $\delta\L$ and $\delta\lcf$ we also need two distinct minimal on-shell bases of operators~$\op_i$ and $\dlcf_i$
\begin{align}
\delta\L=\bop_i(a\mu,\bar{g})Z_{ij}^\op\op_j\,,&&\delta\lcf= \cop_i(a\mu,\bar{g})Z^\lcf Z_{ij}^\dlcf\dlcf_j
\end{align}
with free coefficients $\bop_i$ and $\cop_i$, renormalised coupling $\bar{g}=\bar{g}(\mu)$ and renormalisation scale $\mu$. Furthermore we introduced the mixing matrices $Z^\op$ and $Z^\dlcf$ satisfying
\begin{equation}
\R[i]\op=Z^\op_{ij}\op_j\,,\quad\R[i]\dlcf=Z^\lcf Z_{ij}^\dlcf\dlcf_j\,,\quad\R\lcf=Z^\lcf\lcf\,,
\end{equation}
where, for later use, $Z^\dlcf$ is defined with the renormalisation factor $Z^\lcf$ of the continuum operator~$\lcf$ divided out, assuming, for simplicity, no mixing for $\lcf$ itself.

Which operators are actually allowed to contribute is determined by the symmetries of the lattice quantity, i.e. for the lattice Yang Mills action we only allow operators complying with
\begin{itemize}
\setlength{\itemsep}{-2pt}
\item Local SU(N) gauge symmetry,
\item C-, P- and T-symmetry,
\item discrete rotation and translation invariance (at least in manifolds without boundaries), i.e. the O(4) symmetry of $\L$ is partially broken and reduced to hypercubic symmetry.
\end{itemize}
With these restrictions and the requirement of linear independence only the following operators are left~\cite{Luscher:1985}
\begin{equation}
\op_1=\frac{1}{g_0^2}\tr(D_\mu F_{\nu\rho}D_\mu F_{\nu\rho})\,,\ \ \op_2=\frac{1}{g_0^2}\sum\limits_\mu\tr(D_\mu F_{\mu\rho}D_\mu F_{\mu\rho})\,,\ \ 
\op_3=\frac{1}{g_0^2}\tr(D_\mu F_{\mu\nu}D_\rho F_{\rho\nu})\EOM 0\,.
\end{equation}
As indicated, the operator $\op_3$ can be dropped from the basis as well since we are only interested in a minimal on-shell basis and thus can use equations of motion~(EOM)~\cite{Luscher:1985,Georgi:1991ch}, i.e. $D_\mu F_{\mu\nu}=0$, for the reduction of the basis.

In analogy to~\cite{Balog:2009yj,Balog:2009np} we can write, as an example, the lattice artifacts of a 
renormalised connected 2-point function as
\begin{eqnarray}
\frac{\lattExp[]{\lcfrlat(x)\lcfrlat(y)}}{\contExp[]{\lcfrlat(x)\lcfrlat(y)}}&=&1+a^2\Bigg(\sum_i \cop_i(1,\bar{g})\delta_i^{\dlcf}(x,y;1/a)-\sum_{j=1}^2\bop_j(1,\bar{g})\delta_j^\op(x,y;1/a)\hspace{-3pt}\Bigg)+\ord(a^4),\quad\quad
\end{eqnarray}
where $\lattExp[]\ldots$ and $\contExp[]\ldots$ are the expectation values in the lattice theory at lattice spacing $a$ and zero respectively.
In the continuum theory we introduce the shorthand
\begin{eqnarray}
{\delta_i^{\dlcf}(x,y;\mu)}&=&\left.\frac{\Exp{\lcfr(x)\dlcfr i(y)}_\mathrm{con}+\Exp{\dlcfr i(x)\lcfr(y)}_\mathrm{con}}{\Exp{\lcfr(x)\lcfr(y)}_\mathrm{con}}\right|_{\mu}
,\\
{\delta_j^\op(x,y;\mu)}&=&\int\D^4z\,\left.\frac{\Exp{\lcfr(x)\lcfr(y)\op_{j;\mathrm{R}}(z)}_\mathrm{con}}{\Exp{\lcfr(x)\lcfr(y)}_\mathrm{con}}\right|_{\mu}
\end{eqnarray}
and choose the renormalisation scale as $\mu=1/a$ when matching our Symanzik effective theory to the lattice theory. As an example we choose on the lattice the regulator independent~(RI) renormalisation condition (with scale $\mu_\RI^{-1}=|\xRI|$)~\cite{Gimenez:2004me}
\begin{align}
\lattExp[]{\lcf_\RI(\xRI)\lcf_\RI(0)}=\lattExp[]{\lcf(\xRI)\lcf(0)}\Big|_{\substack{g_0=0\\a=0\hfill}}=\Xi_\mathrm{RI,lat}(\bar{g}_\mathrm{lat}(1/a);a)\lattExp[]{\lcfrlat(\xRI)\lcfrlat(0)}.
\end{align}
The equation defines $\Xi_{\RI,\mathrm{lat}}$ which relates the MS lattice to our RI scheme in terms of
the correlator at a specific distance. % and is treated here like an observable.
For this choice we find for the leading lattice artifacts as $a\searrow0$
\begin{align}
\frac{\lattExp[]{\lcf_\RI(x)\lcf_\RI(y)}}{\contExp[]{\lcf_\RI(x)\lcf_\RI(y)}}={}&1+a^2\Bigg(\sum\limits_i\TL \cop_i\Big\{\delta_i^{\dlcf}(x,y;1/a)-\delta_i^{\dlcf}(\xRI,0;1/a)\Big\}\label{eq:LeadingLattArtifacts}\\
-\sum\limits_{j=1}^2&\TL \bop_j\Big\{\delta_j^\op(x,y;1/a)-\delta_j^\op(\xRI,0;1/a)\Big\}\Bigg)\times[1+\ord(\bar{g}^2({1/a}))]+\ord(a^4)\nonumber,
\end{align}
where $\TL \cop_i$ and $\TL \bop_j$ are the tree-level values of the coefficients and can be obtained from a simple Taylor-expansion in $a$ in the classical theory.
Note that if a coefficient vanishes at tree-level the 1-loop coefficient might be needed to obtain the leading powers in the coupling.

\section{Lattice artifacts and the Renormalisation Group}\label{sec:RG}
In order to understand in which way the $\delta_i^\dlcf(x,y;1/a)$ and $\delta_j^\op(x,y;1/a)$ have an impact on pure $a^2$~scaling we need to understand how they react to a change in the lattice spacing or equivalently to a change of the renormalisation scale $\mu$.
This is governed by the Renormalisation Group equations
\begin{align}
\mu\frac{\D\delta_i^\dummy(x,y;\mu)}{\D\mu}=\gamma_{ik}^\dummy\delta_k^\dummy(x,y;\mu),\quad\dummy\in\{\op,\dlcf\},
\end{align}
with the anomalous dimension matrices
\begin{align}
\gamma^\dummy_{ik}=\mu\frac{\D Z^\dummy_{ij}}{\D\mu}(Z^\dummy)^{-1}_{jk}=-(\TL\gamma_\dummy)_{ik}\bar{g}^2(\mu)+\ord(\bar{g}^4(\mu))
\end{align}
and the corresponding mixing matrices $Z^\dummy$.
We choose now the operator bases such that $\TL\gamma_\dummy=\diag\lbrace\TL\gamma_{\dummy,1},\ldots,\TL\gamma_{\dummy,n}\rbrace$ and introduce the Renormalisation Group Invariants~(RGI)
\begin{equation}
D_j^\dummy(x,y)=\lim_{\mu\rightarrow\infty}\left[2b_0\bar{g}^2(\mu)\right]^{-\hat{\gamma}_j^\dummy}\delta_j^\dummy(x,y;\mu)\,,\quad\hat\gamma_j^\dummy=\frac{\TL\gamma_{\dummy,j}}{2b_0}\,,
\end{equation}
where $b_0$ is the 1-loop coefficient of the $\beta$-function.
This allows to rewrite
\begin{eqnarray}
\delta_i^\dummy(x,y;\mu)&=&\left[2b_0\bar{g}^2(\mu)\right]^{\hat{\gamma}_i^\dummy}\Pexp\left[\int\limits_0^{\bar{g}(\mu)}\D u\left\{\frac{\gamma^\dummy(u)}{\beta(u)}-\frac{2\hat\gamma^\dummy}{u}\right\}\right]_{ij}D_j^\dummy(x,y)\\
&=&\left[2b_0\bar{g}^2(\mu)\right]^{\hat\gamma_i^\dummy}D_i^\dummy(x,y)+\ord\left(\left[\bar{g}^2(\mu)\right]^{1+\hat\gamma_i^\dummy}\right),
\end{eqnarray}
where $\Pexp$ denotes the path ordered exponential with decreasing $\bar{g}$ from the left to the right.
As a result all dependence on the renormalisation scale is absorbed into the prefactor of the RGI.
Inserting this into equation~\eqref{eq:LeadingLattArtifacts} yields the desired form
\begin{align}\label{eq:LeadingLattArtifactsRGI}
\frac{\lattExp[]{\lcf_\RI(x)\lcf_\RI(y)}}{\contExp[]{\lcf_\RI(x)\lcf_\RI(y)}}&=1+a^2\Bigg(\sum\limits_i\TL \cop_i[2b_0\bar{g}^2(1/a)]^{\hat{\gamma}_i^\dlcf}\Big\{D_i^{\dlcf}(x,y)-D_i^{\dlcf}(\xRI,0)\Big\}\\
-\sum\limits_{j=1}^2&\TL \bop_j[2b_0\bar{g}^2(1/a)]^{\hat{\gamma}_j^\op}\Big\{D_j^\op(x,y)-D_j^\op(\xRI,0)\Big\}\Bigg)\times [1+\ord(\bar{g}^2({1/a}))]+\ord(a^4).\nonumber
\end{align}
So far we included contributions of the field~$\lcf$ to the leading lattice artifacts.
Since they depend on the fields of interest, we now restrict ourselves to the contributions from the action. This is sufficient for spectral quantities, such as masses and energies.
Such quantities depend only on the quantum
numbers of the fields from which they are extracted.

\section{Computing 1-loop anomalous dimensions}\label{sec:CompAnomDim}
In equation~\eqref{eq:LeadingLattArtifactsRGI} we need the leading anomalous dimensions of the operators $\op_i$. We employed two strategies extracting these anomalous dimensions from different types of Euclidean Green's functions:\vspace{-5pt}
\begin{enumerate}
  \setlength{\parskip}{1pt}
  \setlength{\itemsep}{1.5pt plus 0.3ex}
\item[(i)] connected on-shell Green's functions of fundamental gauge fields~$A$ with insertion of an operator~$\tilde\op[A](q)$ carrying an overall momentum~$q\neq0$, see e.g.~\cite{Gracey:2002he},
\begin{eqnarray}
G_{i}^{ab}(p_1,p_2;q)&=&\Exp{\eta_1\cdot\R{\tilde{A}}^a(p_1)\;\eta_2\cdot\R{\tilde{A}}^b(p_2)\;\tilde\op_i[A](q)}_\mathrm{con}\\
G_{i}^{abc}(p_1,p_2,p_3;q)&=&\Exp{\eta_1\cdot\R{\tilde{A}}^a(p_1)\;\eta_2\cdot\R{\tilde{A}}^b(p_2)\;\eta_3\cdot\R{\tilde{A}}^c(p_3)\;\tilde\op_i[A](q)}_\mathrm{con}
\end{eqnarray}
with external momenta~$p_j$ and polarisation vectors~$\eta_j$ fulfilling the (Minkowskian) on-shell condition
\begin{equation}
(p_j)_0^2=-\vec{p}_j^2,\quad p_j\cdot\eta_j=0\,.
\end{equation}
\item[(ii)] 1PI off-shell Green's functions of classical background fields~$B$ with insertion of an operator~$\tilde\op[B+A](q)$ using the background field~(BGF) method~\cite{tHooft:1975uxh,Abbott:1980hw}%,Abbott:1981ke,Luscher:1995vs}
\begin{eqnarray}
\Gamma_{i}^{ab}(p_1,p_2;q)&=&\Exp{\eta_1\cdot\tilde{B}^a(p_1)\;\eta_2\cdot\tilde{B}^b(p_2)\;\tilde\op_i[B+A](q)}_\mathrm{1PI}\\
\Gamma_{i}^{abc}(p_1,p_2,p_3;q)&=&\Exp{\eta_1\cdot\tilde{B}^a(p_1)\;\eta_2\cdot\tilde{B}^b(p_2)\;\eta_3\cdot\tilde{B}^c(p_3)\tilde\op_i[B+A](q)}_\mathrm{1PI}
\end{eqnarray}
with $q=0$ but no additional constraints on the external momenta~$p_j$.
\end{enumerate}
Both strategies allow us to restrict considerations to 2- and 3-point functions as well as 1-loop computations.
However this results in an enlarged set of operators during renormalisation by introducing additional redundant operators namely total divergence operators 
\begin{equation}
\opT_1=\frac{1}{g_0^2}\partial_\mu\tr(F_{\nu\rho}D_\mu F_{\nu\rho})\,,\quad\opT_2=\frac{1}{g_0^2}\sum_\mu\partial_\mu\tr(F_{\mu\rho}D_\mu F_{\mu\rho})\,,
\end{equation}
for case~(i) and the previously discarded EOM vanishing operator $\E_1 = \op_3$ for case~(ii).

In the presence of either set of redundant operators we can extract the desired on-shell mixing matrix $Z^\op$ from
\def\red{Q}
\begin{equation}
\R{\begin{pmatrix}
\op_i\\[6pt]
\red_k
\end{pmatrix}}
=\begin{pmatrix}
Z^\op_{ij}&Z^{\op \red}_{il}\\[6pt]
0&Z^{\red}_{kl}
\end{pmatrix}\begin{pmatrix}
\op_j\\[6pt]
\red_l
\end{pmatrix}\,,\quad \red\in\{\opT,\E\}\,,
\end{equation}
where the triangular mixing matrix ensures that the renormalised $\red$ vanish for operator momentum $q=0$ ($\red=\opT$) and on-shell quantities ($\red=\E$).
Due to this structure we only need tree-level contributions from the redundant operators to determine $Z^{\op\red}$ in order to extract $Z^\op$ at 1-loop.

Being only interested in 1-loop anomalous dimensions, another simplification can be achieved by use of the identity, see e.g.~\cite{Misiak:1994zw,Chetyrkin:1997fm},
\begin{equation}
\frac{1}{(k+p)^2}=\frac{1}{k^2+\Omega}-\frac{2kp+p^2-\Omega}{[k^2+\Omega](k+p)^2}\,,
\end{equation}
where $k$ is the loop momentum, $p$ a constant vector and $\Omega>0$.
Since the second term on the right hand side is one power less UV divergent we can iterate this step until the terms of the form ${[k^2+\Omega]^{-n}}$ carry all the UV divergences.
The leftover terms are finite or at most IR divergent and can be dropped because only UV divergences contribute to anomalous dimensions in mass-independent renormalisation schemes such as $\overline{\mathrm{MS}}$ that we employ in $D=4-2\epsilon$ dimensions. Consequently only the basic integrals of dimensional regularisation $\int\mathrm{d}^Dk\,(k^2)^m/[k^2+\Omega]^n$
are left to evaluate.

The Feynman graphs contributing to the 2- and 3-point functions were obtained using QGRAF \cite{Nogueira:1993,Nogueira:2006pq}.
The generation of Feynman rules for the operators and the computational steps of dimensional regularisation were then performed in FORM~\cite{Vermaseren:2000nd}.

Independently of the strategy we find\footnote{For more details see~\cite{H:inprep}.} for the 1-loop mixing matrix $Z^\op$ in the $\overline{\mathrm{MS}}$ scheme
\begin{equation}
\R{\begin{pmatrix}
\op_1\\[6pt]
\op_2
\end{pmatrix}}=
\begin{pmatrix}
1+\frac{7\CA}{3\epsilon}
 \frac{\bar{g}^2}{(4\pi)^2}&0\\[3pt]
-\frac{7\CA}{15\epsilon}
 \frac{\bar{g}^2}{(4\pi)^2}&1+\frac{21\CA}{5\epsilon}
 \frac{\bar{g}^2}{(4\pi)^2}
\end{pmatrix}
\begin{pmatrix}
\op_1\\[6pt]
\op_2
\end{pmatrix}\,.
\end{equation}
Since this basis is not diagonal we switch to the diagonal basis
\begin{align}
\basis_1=\op_1\,,\quad\hat{\gamma}_1^\basis=\frac{7}{11}\approx0.636&&\basis_2=\op_2-\frac{1}{4}\op_1\,,\quad\hat{\gamma}_2^\basis=\frac{63}{55}\approx 1.145\label{eq:anomDimRes}\,.%\\
%\,,&\,.\nonumber
\end{align}
Note that the first anomalous dimension has been known for quite some time~\cite{Narison:1983}.

\def\mass{m^\Phi}
\def\vecx{\mathbf{x}}
\section{Conclusion}
Coming back to our example of the 2-point function we find for the lowest contributing mass~$\mass$, i.e. the lightest glueball mass~$\mass$ with the quantum numbers of $\lcf$,
\begin{align}
\mass_{\mathrm{lattice}}={}&-\lim_{x_0\rightarrow\infty}\frac1a\ln\frac{\sum\limits_\vecx\lattExp[]{\lcf_\RI(x_0+a\hat{0},\vecx)\lcf_\RI(0)}}{\sum\limits_\vecx\,\lattExp[]{\lcf_\RI(x_0,\vecx)\lcf_\RI(0)}}\\
\approx{}&\mass_\mathrm{cont}-\frac{a^2}{2}\left(\TL{\bbase}_1[2b_0\bar{g}^2(1/a)]^{0.636}\Delta_1+\TL{\bbase}_2[2b_0\bar{g}^2(1/a)]^{1.145}\Delta_2+\ldots\right)+\ord({a^4})\nonumber
\end{align}
where $\Delta_i=\langle \lcf_0|\RGI[i;]{\basis}(0)|\lcf_0\rangle$ with zero-momentum ground state $|\lcf_0\rangle$ and $\TL{\bbase}_i$ are the tree-level coefficients of the diagonalised basis, e.g. $\TL{\bbase}_1=1/48$, $\TL{\bbase}_2=1/12$ for the plaquette action~\cite{Luscher:1985}.
As expected no dependence on the definition of $\lcf$ or the chosen renormalisation condition remains for the spectral quantity.

More generally, since $\bar{g}^2(1/a)\sim-1/\ln(a\Lambda)$ and $\hat{\gamma}^\basis_i>0$, the leading logarithmic corrections originating from the action accelerate the convergence of the continuum limit.
With $\hat\gamma_1^\basis<\hat\gamma_2^\basis$ the leading contribution comes from the O(4) symmetric operator $\basis_1$.
In case of Symanzik improved actions one replaces $\bbase_i^{(0)}\rightarrow \bbase_i^{(n)}[\bar{g}^2(1/a)]^{n}$, where $n=1$ means tree-level improvement.

This information suffices to back up the usually employed naive $a^2$ extrapolations in lattice Yang Mills theory for all spectral quantities.
However, other quantities need additional information namely the matching coefficients and leading anomalous dimensions for the $\dlcf_i$.
Especially the new sector of Gradient flow observables with flow time $t$, requires the inclusion of an additional operator, located at the boundary $t=0$, in the Symanzik effective action
of the 5-dimensional formulation~\cite{Ramos:2015baa}.

\input{Proceedings_Husung.bbl}

\end{document}

%% file: Proceedings_Husung.bbl
\providecommand{\href}[2]{#2}\begingroup\raggedright\endgroup